\documentclass[10pt,conference]{IEEEtran}
\IEEEoverridecommandlockouts
\usepackage{cite}
\usepackage{amsmath,amssymb,amsfonts}
\usepackage{algorithmic}
\usepackage{tabularx}
\usepackage{graphicx}
\usepackage{textcomp}
\usepackage{subcaption}

\usepackage{tablefootnote}
\usepackage[flushleft]{threeparttable}
\usepackage{makecell}
\usepackage{tikz}
\usepackage{xcolor}
\def\BibTeX{{\rm B\kern-.05em{\sc i\kern-.025em b}\kern-.08em
    T\kern-.1667em\lower.7ex\hbox{E}\kern-.125emX}}

\usepackage{tabularx}
\usepackage{multirow}
\usepackage{booktabs} %
\usepackage{caption}
\usepackage{placeins}
\usepackage{url}
\usepackage{framed}

\usepackage{amsmath}

\usepackage{tikz}
\usepackage{pgfplots}
\pgfplotsset{width=10cm,compat=1.9}
\usepackage{xcolor}
\usetikzlibrary{arrows.meta,
                chains,
                positioning}

\usepackage{verbatim}
\usepackage{xspace}
\usepackage{hyperref}

\usepackage[breakable, theorems, skins]{tcolorbox}
\tcbset{enhanced}

\DeclareRobustCommand{\mybox}[2][gray!30]{%
\begin{tcolorbox}[   %
        breakable,
        left=0pt,
        right=0pt,
        top=0pt,
        bottom=0pt,
        colback=#1,
        colframe=#1,
        width=\dimexpr\columnwidth\relax, 
        enlarge left by=0mm,
        boxsep=5pt,
        arc=0pt,outer arc=0pt,
        ]
        #2
\end{tcolorbox}
}

\usepackage{svg}

\definecolor{darkgreen}{rgb}{0.05,0.5,0.05}

\newcommand{\OSSFSG}{Open Source Software for Social Good\xspace}

\newcommand{\gh}{GitHub\xspace}

\usepackage{framed}
\usepackage[strict]{changepage}

\definecolor{PAblue}{RGB}{0,122,204}%

\newcommand{\quo}[2]{ 
	\vspace{-0.15cm}
	\def\FrameCommand{%
		\hspace{0pt}%
		{\color{PAblue}\vrule width 2pt}%
		{\color{white}\vrule width 2pt}%
		\colorbox{white}
	}%
	\MakeFramed{\advance\hsize-\width\FrameRestore}%
	\noindent\hspace{-4.55pt}%
	\begin{adjustwidth}{}{0pt}
		\vspace{-2pt}%
		``\emph{#1}'' ({#2})
		\vspace{-3pt}
	\end{adjustwidth}\endMakeFramed%
	\vspace{-0.15cm}
}

\newcommand{\inlinequo}[2]{``\emph{#1}'' (#2)}

\begin{document}
\bstctlcite{IEEEexample:BSTcontrol}

\title{Leaving My Fingerprints: Motivations and Challenges of Contributing to OSS for Social Good}

\author{\IEEEauthorblockN{Yu Huang}
\IEEEauthorblockA{\textit{University of Michigan} \\
Ann Arbor, MI \\
yhhy@umich.edu}
\and
\IEEEauthorblockN{Denae Ford}
\IEEEauthorblockA{\textit{Microsoft Research} \\
Redmond, WA USA \\
denae@microsoft.com}
\and
\IEEEauthorblockN{Thomas Zimmermann}
\IEEEauthorblockA{\textit{Microsoft Research} \\
Redmond, WA USA \\
tzimmer@microsoft.com}}

\maketitle

\begin{abstract}
 
 When inspiring software developers to contribute to open source software, the act is often referenced as an opportunity to build tools to support the developer community. However, that is not the only charge that propels contributions---growing interest in open source has also been attributed to software developers deciding to use their technical skills to benefit a common societal good. To understand how developers identify these projects, their motivations for contributing, and challenges they face, we conducted 21 semi-structured interviews with OSS for Social Good (OSS4SG) contributors. From our interview analysis, we identified themes of contribution styles that we wanted to understand at scale by deploying a survey to over 5765 OSS and \OSSFSG contributors. From our quantitative analysis of 517 responses, we find that the majority of contributors demonstrate a distinction between OSS4SG and OSS. Likewise, contributors described definitions based on \emph{what} societal issue the project was to mitigate and \emph{who} the outcomes of the project were going to benefit. In addition, we find that OSS4SG contributors focus less on benefiting themselves by padding their resume with new technology skills and are more interested in leaving their mark on society at statistically significant levels. We also find that OSS4SG contributors evaluate the owners of the project significantly more than OSS contributors. These findings inform implications to help contributors identify high societal impact projects, help project maintainers reduce barriers to entry, and help organizations understand why contributors are drawn to these projects to sustain active participation.
\end{abstract}

\maketitle

\section{Introduction}
Open source software (OSS) has enhanced the tools developers build and likewise has inspired developers to consider how the tools they build can support the greater good~\cite{boss2012coding, wagener2017how}.
However, when we refer to OSS we often consider the projects that are building developer tools~\cite{hertel2003motivation}, or extending libraries to popular programming language projects in Python like TensorFlow~\cite{tensorflow} or Javascript like React~\cite{react}. 
Although contributions to these types of OSS projects are valuable to the community, there are other galvanizing projects with broader impacts that inspire developers to contribute to OSS. Specifically, projects that can directly benefit society: Projects such as \textit{Little Window}~\cite{littlewindow}, which supports domestic violence victims with resources to leave abusive relationships, \textit{RestroomRefuge}~\cite{restroom}, which helps identify safe restroom access for transgender and gender nonconforming individuals, 
or \textit{CommCare}~\cite{dimagi}, which is widely used during the COVID-19 pandemic to support frontline workers in developing countries, seem to be fewer in quantity but powerful in influence.
In that same vein, there is an emerging number of software companies which are traditionally profit-driven by building software for a technological good but are now partnering with non-profits to promote projects to tackle societal issues and improve lives~\cite{ms-ai4good, ai4earth, ai4good}.

While all of these projects seem to present non-traditional software categories, the common property they share is to support a greater \textit{social good}. 
Though without a converged definition, the term \textit{social good} has been introduced from domains like sociology~\cite{mor2019social,mor2020practice} to Artificial Intelligence~\cite{abebe2018mechanism} and software engineering~\cite{ferrario2014software} in recent years, but is still a new concept for the open source community. 
In this paper, to the best of our knowledge, we for the first time introduce the notation of \textit{\textbf{\OSSFSG (OSS4SG)}} to refer to these projects above that benefit society.
Although these projects are rarer in open source, they still manage to find contributors that are deeply motivated to offer their expertise. For instance, the \textit{CommCare} project has been supported by open source developers during the COVID-19 pandemic, as it previously did during the Ebola outbreak, with case reporting, contact tracing, and community education~\cite{covid}. 
But is this excitement to contribute to social good projects any different from the motivations to contribute to other OSS projects~\cite{lee2017understanding}?
Prior research in social work presents theoretical underpinnings of how social good movements can be fueled by a sense of urgency for broad societal changes---bringing in new  perspectives along the way~\cite{mor2019social}. 
Likewise, in the computer science education field projects with broader societal impacts can also attract a broader range of participation across minority groups~\cite{payton2016stars}.
Is that also the case for OSS4SG?
From \gh 's Social Impact Report, we identify that there are several non-profits that are also interested stake holders in using OSS4SG to make a social impact~\cite{github2020socialsectorreport}, but ultimately little is known about the contributors in this community.
As the first study to understand the full range of contribution interest and challenges in OSS4SG, we investigate how OSS4SG contributors are different from general OSS contributors. In this paper, we sought out to understand 
a) differences between OSS and OSS4SG projects from contributors' perspectives, 
b) why contributors decide to contribute to OSS4SG, 
c) how contributors select which OSS4SG project to work on,
and d) what challenges developers face when contributing in OSS4SG.

To understand the scope of identifiable OSS4SG projects, we first identified two public third-party software sites, Ovio~\cite{ovio} and Digital Public Goods~\cite{digitalpublicgoods}, as two of the few curated repositories of nominated OSS4SG projects.
We took an exploratory sequential mixed-methods approach~\cite{creswell2018research} where we first, conducted semi-structured interviews with 21 active OSS4SG contributors from around the world and second, used the 17 hours of analyzed interview data to deploy a quantitative survey to understand differences in project types from both OSS4SG and general OSS contributors.
From our interviews, we find contributors working on interesting projects ranging from domestic violence support to finding safe restrooms.
From our interview participants we find that there is a distinction in defining OSS4SG as ``It's complicated because part of the point comes down to outcome.''
Similarly, in our survey we find that over 52\% of respondents felt similarly and regarded OSS4SG to be distinct from OSS. 
Using the combined results from our study and prior literature~\cite{ferrario2014software}, in this study, we define \textbf{\textit{open source for social good}} as follows:
\mybox{Open source software projects where the outcome distinctly targets a community of people to overcome a societal issue.} \noindent We should note that this definitions considers both the societal issue and \emph{who} the project intends to impact.

We also find that, while sharing common features with OSS, OSS4SG holds unique characteristics. Compared to OSS, OSS4SG contributors care significantly less about benefiting themselves including learning skills and improving portfolios, but emphasize significantly more on addressing societal issues. 
OSS4SG contributors also evaluate the trust of owners of projects with significantly more care and are specifically facing with more challenges of matching themselves with projects.

The main contributions from this paper are:
\begin{enumerate}
    \item A contributor informed definition of \OSSFSG. This definition also takes into consideration of how non-OSS4SG contributors describe social impact.

    \item A taxonomy and investigation of motivations, projects identification signals, and challenges for \OSSFSG projects and how they vary from general OSS contributions.
    \item Design recommendations for project maintainers and organizations for how to support OSS contributors who are interested in projects with societal impacts such as having an impact statement on contributing guidelines.
\end{enumerate}

\section{Methodology}
To explore how open source contributors work on projects with societal impacts, we conducted a mixed-methods study of semi-structured interviews of OSS4SG contributors and distributed a survey to a range of OSS contributors. 
The ethics for this study were reviewed and approved by the Microsoft Research Institutional Review Board (MSRIRB), which is an IRB federally registered with the United States Department of Health \& Human Services. (Reference: MSRIRB \#649 and \#718).
All study materials can be found %
online~\cite{suppmaterials} 
\href{https://doi.org/10.5281/zenodo.4536791}{\includegraphics[width=3cm,trim=0 1mm 0 0]{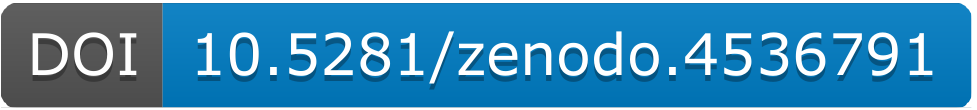}}.

\subsection{Research Questions}
We investigated the following research questions:
\begin{description}
\setlength{\itemsep}{0pt}
\setlength{\parskip}{0pt}
    \item[\textbf{RQ1}] {How do contributors define OSS for Social Good?} %
    
    \item[\textbf{RQ2}] {What motivations do contributors have to contribute to OSS for Social Good?}  %
    \item[\textbf{RQ3}] {What factors do contributors consider to select an OSS for Social Good project?} %
    \item[\textbf{RQ4}] {What are the current challenges to work in OSS for Social Good?} %
\end{description}

\subsection{Semi-Structured Interviews}
To understand the experiences of OSS4SG contributors we conducted interviews. Since there has been no research on open source for social good, we first had to identify opens source for social good projects before recruiting participants.

\subsubsection*{\textbf{Identifying Open Source for Social Good Projects}}
\label{sec:SG-selection}
As there are no lists of projects currently hosted by large open source platforms, we reviewed \gh's Social Impact report from for reliable sources~\cite{github2020socialsectorreport}.
From there, we identified two public third-party software project resources: \textit{Ovio}~\cite{ovio} and \textit{Digital Public Goods}~\cite{digitalpublicgoods}, which provide a nomination-based list of software projects publicly labeled with social good topics. 
Ovio uses seven social labels\footnote{\textit{Ovio}'s labels for social good: Civil Tech, Economics, Education, Environment \& Nature, Health \& Well-being, Humanitarian, Society \& Culture.} 
and \textit{Digital Public Goods} adopts the 17 sustainable development goals (SDGs) from the United Nations~\cite{sdg} as the social good labels for their projects. %
We sample OSS4SG interview participants and survey respondents from these curated lists.

\subsubsection*{\textbf{Participants}} 
We recruited 21 interview participants from 437 OSS4SG projects where a valid link to a \gh project was available.  (1) We purposely sampled 80 projects across 4 quartiles of repository stars (e.g., an indicator of community interest). (2) Our recruitment strategy to select participants from the selected projects had a two-pronged approach: first, we emailed authors of the 60 most recent commits as of June 1, 2020 and second, we picked one contributor randomly from each quartile of that projects commit history. In total we emailed 250 contributors and had response rate of 8.4\%~\footnote{The actual response rate is higher because more contributors contacted us to participate in the interviews but there were no slots available.}. 
The repository links and the labels of corresponding social good topics for all of the 437 projects are available online~\cite{suppmaterials}.

Table~\ref{tab:participants} lists demographic information of our interview participants including their gender, the country they contribute from, and the reported project domains. We also collected information on race and ethnicity which can be found on our project website.
Among our participants, P21 identified themself as a Senior Technical Advisor for their project.  They have a background in Pharmaceuticals and was added to the contributor team of a social good project on health management science as a supervisor, but did not personally participate in any open source activities. In the end of this paper, we will discuss separately on P21's observations and comments from the perspective of technical advisors (at the end of Section~\ref{sec:RQ4}). Otherwise, all the reported results of the interviews are from P1 to P20 who identify as contributors. The interviews reached theoretical saturation at P20. 
Participants reported 3.05 years of experience on OSS4SG on average ($sd = 2.98$) and 4.38 years of experience on OSS in general ($sd = 4.07$).

\begin{table}
    
    \centering
    \caption{Interview Participants \label{tab:participants}\vspace{-.1cm}}\small
    \begin{tabular}{@{}l@{\phantom{x}}cc@{\phantom{x}}cc@{}c@{}}\toprule
        \textbf{ID} &\textbf{Gender} & \textbf{\makecell{SG\\ Exp}}& \textbf{\makecell{OSS\\ Exp}} & \textbf{\makecell{Location of \\Contribution}} & \textbf{\makecell{Project\\Domains}} \\
        \midrule
          P1 & W & 2 & 2  & Mexico & Crypto, Security\\[3pt]
          P2 &  M & 1 & 1  & USA & Finance\\[3pt]
         P3 &  M &  8 & 8 &  Germany  & \makecell{Education, Healthcare,\\ Disaster Tracking\vspace{3pt}} \\[3pt]
         P4 & W & 1 & 1 & UK & \makecell{Charity, \\Domestic Violence\vspace{3pt}} \\[3pt]
         P5 & M & 1 & 3 & India & Environment \\[3pt]
         P6 & M & 0.5 & 10 & Turkey & COVID-19 Tracking \\[3pt]
         P7 & M & 0.5 & 0.5 & India & \makecell{Education,\\ Environment\vspace{3pt}} \\[3pt]
         P8 & M & 4 & 5.5 & Israel & \makecell{Structurally-Safe\\ Buildings}\\[3pt]
         P9 & M & 8 & 8 & Australia & Healthcare, Education \\[3pt]
         P10 & W & 2 & 2 & India & Healthcare, Education \\[3pt]
         P11 & W & 0.5 & 0.5 & India & Education \\[3pt]
         P12 & M & 2 & 2 & USA & COVID-19 Tracking \\[3pt]
         P13 & M & 2 & 6 & USA & \makecell{Education, \\Non-profit Tools\vspace{3pt}} \\[3pt]
         P14 & NB & 8 & 8 & Germany & \makecell{Anti-Gentrification,\\ Safe Restrooms\vspace{3pt}} \\[3pt]
         P15 & M & 10 & 10 & Spain & \makecell{eGovernment,\\Civil Participation\vspace{3pt}} \\[3pt]
         P16 & M & 0.5 & 0.5 & India & \makecell{Healthcare\vspace{3pt}} \\[3pt]
         P17 & M & 2 & 2 & India & Education \\[3pt]
         P18 & M & 1.5 & 1.5 & Romania & Local Administration \\[3pt]
         P19 & M & 0.5 & 1 & India & Healthcare \\[3pt]
         P20 & M & 5 & 15 & Canada & \makecell{Management for \\Government and Charity\vspace{3pt}} \\[3pt]
         P21* & M & 5 & 5 & USA & Healthcare\\
         \bottomrule
    \end{tabular}\vspace{-.4cm}
\end{table}

\subsubsection*{\textbf{Protocol}}
We conducted semi-structured interviews remotely ranging from 45-55 minutes. Two of the authors conducted the interviews. %
 The interviews topics included: contributor background, perceptual differences between OSS4SG,  motivations and factors considered when selecting an OSS4SG project, experience comparisons in OSS and OSS4SG, and challenges as well as expectations for OSS4SG.
 
After the interview, each participant was compensated with a \$25 USD equivalent electronic gift card for their participation.
The format of semi-structured interviews are widely used in software engineering studies~\cite{shrestha2020here}, which allows the flexibility to dig into more specific questions based on interviewees' own experience.
All the interviews were recorded and later transcribed by a third-party transcription service. 
The first two interviews were conducted as pilot interviews that included a superset of the questions, based on which we finalized the interview questions for the remaining interviews.

\subsubsection*{\textbf{Analysis}} 
\label{sec:analysis}
We conducted inductive thematic analysis~\cite{huberman2014qualitative} on the interview transcripts over multiple phases using \textit{ATLAS.ti}. We first generated open codes by labelling notable recurring statements made by the participants. The first two authors analyzed 2 transcripts independently to identify open codes and then convened to discuss codes and determine relationships. Once refined the first author continued labeling data based on our iterative and collaborative refinement of our code book. Next, we collaboratively identified relationships between the codes. Finally, we collaboratively organized codes into meaningful themes. The code book used in our analysis is available online~\cite{suppmaterials}.

\subsection{Survey}
To validate and quantify our themes from interviews, we distributed a survey that include both OSS4SG and the broader population of open source contributors on \gh.

\subsubsection*{\textbf{Identifying Open Source \& Social Good Contributors}}
~\label{sec:survey-recruitment}
We recruited \gh contributors from two types of projects.
(1) \textit{Set-OSS4SG: Open Source Projects for Social Good.} This is the same set of 437 \gh projects that are labeled as for social good (see Section~\ref{sec:SG-selection}). We had a population of 14,256 contributors, among which 7,500 contributors' contact information is publicly available.
(2) \textit{Set-OSS: Open Source Projects excluded from Set-OSS4SG.} To select a comparable set of general open source projects to \textit{Set-OSS4SG}, we ended up randomly selecting 642 projects from \gh with no specific filtering applied. We had a population of 17,978 contributors in total, among which 9409 contributors' contact information were publicly available.

\subsubsection*{\textbf{Respondents}}To verify that the distributions of contributors (i.e., team size) in \textit{Set-OSS4SG} and \textit{Set-OSS} are comparable, we conducted a two-tailed T-test ($p=0.31$) and found no significant difference on the distribution of contributors between \textit{Set-OSS4SG} and \textit{Set-OSS}.

Then, we follow a two-step selection process to get 3,000 contributors from each  of \textit{Set-OSS4SG} and \textit{Set-OSS} to make sure we have a variety of perspectives included in our analysis:
\begin{enumerate}
    \item Most recently active contributors: We randomly selected four contributors from every project in the most recent 60 commits.
    \item Additional contributors: Depending on the final list of contributors from Step~1, we then randomly selected a set of contributors (exclusive from contributors selected in Step~1) from the entire contributor pool so that we have a list of 3,000 contributors in total.
\end{enumerate}

Finally, for \textit{Set-OSS4SG}, we ended up with 1436 active contributors (from Step~1) and 1564 randomly selected contributors (from Step~2). For \textit{Set-OSS}, we ended up with 1383 active contributors (from Step~1) and 1617 randomly selected contributors (from Step~2). We sent out the survey invitation to 5765 reachable emails (some emails are valid but not reachable any more). 
The proportion of women respondents from \textit{Set-OSS4SG} ($n=21$) is relatively higher than \textit{Set-OSS} ($n=7$). Likewise, the race/ethnicity distribution of response were also noticeably higher among Black or African-American in \textit{Set-OSS4SG} ($n=14$) than \textit{Set-OSS} ($n=3$).

\subsubsection*{\textbf{Protocol}}
We used our results from the interviews to design our 28-question survey. Respondents completed the survey in about 20 minutes.
Topics covered in our survey included but are not limited to demographics, roles in projects, perceptions of important factors during selection, and challenges faced. 
To remind respondents to answer questions accordingly, we emboldened text regarding differences between their OSS and OSS4SG experiences.
We also include a set of questions about the level of impact desired using a trolley problem~\cite{thomson1985trolley} based approach to understand interest along the three dimensions of the construal-level theory of psychological distance which are spatial, temporal, and social~\cite{trope2010construal}. 
For our demographic questions, we followed the FDA's description of race and ethnicity~\cite{fda2016raceethnicity} and the HCI Guidelines for Gender Equity and Inclusivity~\cite{scheuerman2019hci}. Most survey questions were listed as optional.
To reduce the risks of participants misunderstanding questions, the survey was reviewed by several people who are not co-authors of this paper. The survey was also piloted with smaller groups before distributing the survey widely.
Upon completion of the survey, participants were able enter a sweepstakes to win one of four \$100 USD electronic gift cards.
The full survey is available online~\cite{suppmaterials}.

\subsubsection*{\textbf{Analysis}} 
From our interviews, we found that perceptions of whether OSS4SG and general OSS are different influence how contributors describe their experiences.
Thus, in our survey, we (1) first collect respondent's opinions on the agreement that OSS4SG distinguish from general OSS, and (2) then whether a respondent has worked in OSS4SG projects before. In the survey analysis, we compare between OSS4SG and OSS.  To clarify the results from our analysis, we use the following named buckets in our text and tables to indicate subgroups of all survey respondents based on their self-reported information.

\begin{itemize}
    \item \textit{P-OSS4SG}: Respondents who agree that OSS4SG should be distinguished from general OSS.
    \item \textit{P-OSS}: Respondents who believe all OSS projects are for social good.
    \item \textit{P-OSS4SG+}: Respondents in P-OSS4SG but have also worked on at least one OSS4SG project in the past.
    \item \textit{P-OSS+}: All respondents excluding P-OSS4SG+.
\end{itemize}

We received 517 valid responses from the 5765 email invitations (8.97\% response rate), which aligns with previous studies in software engineering surveys~\cite{shull2007guide} and comparable to other surveys on open source contributors. Every survey respondent was regrouped into either P-OSS or P-OSS4SG based on their responses and overall actual experience. Our respondents were also from a range of regions including Africa ($n=23$), the Americas ($n=181$), Asia ($n=137$), Europe ($n=164$), and Oceania ($n=11$).
The OSS project domains reported in the survey (499 reported topics in total) covered all 17 sustainable development goals (SDGs) defined by the United Nation; the most popular topics are \textit{Quality Education} (78 projects), \textit{Good Health and Well-Being} (62 projects), and \textit{Industry, Innovation, and Infrastructure} (60 projects).

\section{Results}
For each research question, we first introduce the interview observations and analysis, then present the corresponding survey results. In some cases, we have anonymized parts of quotes to maintain participants’ privacy.

\subsection{\textbf{How do contributors define OSS4SG? (RQ1)}} \label{sec:RQ1}
Though the term \textit{social good} has been used in the literature of open source~\cite{ferrario2014software}, no consensus of definition, especially from the perspective of open source developers, has emerged (see Section~\ref{sec:related-work-se}). We first investigate how OSS contributors understand and define \textit{OSS4SG}.

\subsubsection{Interview} 
Our interview participants expressed two high-level opinions on the definition of \textit{OSS4SG}.
Three out of 20 participants (P5, P8, P11) believed, to an extent, that all OSS projects are for social good, because every project can benefit its users: 
\quo{I think the fact that the project is open source and people can contribute their knowledge and can use it as they are required by the license, there's a social benefit that everyone can use from the project itself.}{P8}
However, participants can draw the same conclusion with different reasoning and understanding of the definition of \textit{social good}. For example:
\quo{I think almost all projects should come into social good because some way or another, you never know what the bigger project is.  So I may be using another open source software for my project, my project being something for kids or for education or for some other social purposes, but the open source project I am working on may be just a tech-based project.}{P11}
Though P11 agrees that all open source projects are for social good, it is based on the assumption that every project can potentially benefit society indirectly by being part of another project that solves societal issues. 

The remaining 17 participants (and to some extent P11) believe that certain unique properties make some open source projects distinct from the rest: identified as for \textit{Social Good}. 
There are three main properties the interview participants used to define \textit{Open Source for Social Good (OSS4SG)}:
\begin{itemize}
    \item The project targets people or communities that need help:
    \inlinequo{It’s just for the software that actually makes a difference on the access or on the usability of people who needs some help.}{P20}
    \item The project aims to solve some societal issues or provide social benefits:
    \inlinequo{If we are talking about the social good term, it comes with any project or any initiative that is, like, that aims to provide any help to the social projects.}{P10}
    \item The project is non-profit (or hosted by non-profit owners):
    \inlinequo{The goal is the software itself, not necessarily profiting off the software.}{P9}
\end{itemize}

Participants also provided inspiring insights on the definition of \textit{OSS4SG} 
from the view of the government's responsibility or users' capability:

\quo{I'm guessing it's things that the government should have been doing, but you're going to have someone else doing it.}{P18}
\quo{Or like you’re building for something who has the knowledge and who in turn builds for someone who doesn’t have the knowledge.}{P19}

\subsubsection{Survey}\label{sec:RQ1-survey}
Among our valid survey respondents, 52.8\% ($n=249$) agreed that OSS4SG projects are distinct from other OSS projects.
We built a logistic regression model to investigate the relationship between demographic information and opinions on distinguishing OSS4SG projects. In this model, we included demographic information, such as gender, race, age, locations of contribution, employment status, professional fields, education levels and background, volunteering experience and capacity of OSS contribution. 
This logistic regression model treats ``OSS4SG is different from OSS'' as 1. 
We found that women in our sample tend to believe all OSS projects are for social good ($\textit{Estimate} = -1.9, p<0.05$). Contributors with previous volunteering experience on human support tend to believe OSS4SG is distinct from general OSS ($\textit{Estimate = 1.72}, p<0.01$).

Among the 249 survey respondents who distinguish between OSS4SG and OSS projects, 89.1\% agreed that the identification of \textit{social good} in open source depended on if a project targeted to solve societal issues, while 77.1\% expected an OSS4SG project to be targeted at users who need help. Only 35.3\% believed OSS4SG projects must be non-profit or hosted by non-profit owners.

\mybox{
\noindent
Contributors cite the target audience and societal issues to determine whether a project is truly OSS for Social Good.
}

\begin{table*}[t!]
    \centering
    \caption{Themes of Motivations for Contributing to OSS for Social Good. \label{tab:motivation-theme}\vspace{-.1cm}}\small
    \resizebox{0.9\textwidth}{!}{%
    \begin{tabular}{@{}p{3.5cm}p{5.5cm}p{5.5cm}p{2.5cm}@{}}\toprule
        \textbf{Theme} & \textbf{Description} & \textbf{Representative Example} & \textbf{Participants}\\ \midrule
        
            To help those in need & Contributors wanted to help people who are in need but may lack the capability of solving the problems themselves. & \em ``I'm so much more motivated to build products that I know have a good outcome for a group of people that is generally underserved.''& P2, P3, P4, P5, P6, \newline P7, P8, P9, P10, \newline P12, P14, P18, P19\\ %
            \midrule
          
          To become a better \newline programmer & Contributors wanted to improve their skills, build up their portfolios, or improve their reputation in the community. &  \em ``when I contribute to that, it can definitely give me more experience.'' & P2, P3, P5, P10, \newline P11, P12, P14, \newline P16, P17, P20\\
          \midrule
          
          To have an impact on \newline society & Contributors wanted to make a difference to the society. & \em ``So, I think the main reason is because I want to make a difference on my life... make a fingerprint on the world.'' & P1, P3, P4, P7, \newline  P13, P14, P15, P17\\ %
          \midrule
          
          For emotional fulfillment & Contributors were motivated by feeling good about the impacts of the project. & \em ``It gives a mental satisfaction that I’m working towards something good''  & {P3, P4, P10, P11, \newline P12, P17, P20} \\
          \midrule

          To help fellow developers with their project & Contributors want to help the developers to achieve the accomplishment of the projects. & \em ``Another is to help the people in the project to help reach their goals.''  & P3, P7, P10, P12, \newline P13, P18\\
          \midrule

          To give back as \newline I received   &  Contributors want to give back to the society (e.g., altruism).&  \em ``And I also feel like however much you take from something, you should give back.'' & P4, P5, P9, P16, \newline P20\\
          \midrule

          To meet like-minded \newline people & Contributors wanted to get to know more people. & \em ``I think it brings like-minded people together most of the time, so I get to interact with people who are working on similar project or they have similar interests.''& P11, P13, P17\\
          \midrule

            As a hobby & Contributors worked in OSS4SG as a hobby or something they like doing.  & \em ``I've moved to sales but still collaborating ... It's just as a hobby.''  &  P14, P15\\ 
          \midrule

          Because I need it \newline for work &  Contributors worked on OSS4SG for their professional work projects. & \em ``So the direct cause that I found it is through [elided]'s little competition.'' & P2\\ %
          
         \bottomrule
    \end{tabular}%
    }%
\end{table*}

\begin{table*}[t]\centering

\caption{The responses from \textit{P-OSS4SG} and \textit{P-OSS} to the question \emph{``Please rate how much you are motivated from the following aspects  when you decide to contribute to a project:''} (Q19, Q19'). 
The columns \textit{P-OSS4SG} and \textit{P-OSS} list the accumulated percentages of responses from ``Important'' and ``Very Important'' (i.e., higher than ``Moderately Important''). \textit{Delta} lists the difference of the percentages between \textit{P-OSS4SG} and \textit{P-OSS}; statistically significant differences are indicated with asterisks (*). The motivation items are ranked and numbered by the importance in \textit{P-OSS4SG}.}
\label{tab:motivation}

  \begin{threeparttable}
\newlength{\myboxheight}
\settoheight{\myboxheight}{1234567890\%}

\def\mybarchart#1{
\resizebox {#1} {\myboxheight} {%
\begin{tikzpicture}[]
\definecolor{clr1}{RGB}{99,99,99}
\definecolor{clr2}{RGB}{240,240,240}
\begin{axis}[
      axis background/.style={fill=gray!10, draw=gray!50},
      axis line style={draw=none},
      tick style={draw=none},
      ytick=\empty,
      xtick=\empty,
      ymin=0, ymax=1, %
      xmin=0, xmax=1]
\addplot [
      ybar interval=.5,
      fill=black,
      draw=none,
]
	coordinates {(1,1) (1,1)}; 
\addplot [
      ybar interval=.5,
      fill=black,
      draw=none,
]
	coordinates {(1,1) (0,1)}; 
\end{axis}%
\end{tikzpicture}%
}%
}

\begin{tabular}{rp{2cm}p{2cm}r}
\toprule
\multicolumn{1}{c}{Motivation}& \multicolumn{1}{l}{\makecell{P-OSS4SG\\ (n=222)}} & \multicolumn{1}{l}{\makecell{P-OSS \\(n=198)}} & \makecell{Delta\\ (P-OSS4SG - P-OSS)}  \\
\midrule
I want to help the target users. (M1) & \mybarchart{23.1pt} 69.4\% & \mybarchart{21.9pt} 65.7\% & 3.7\%\phantom{ ***}  \\ 
I want to give back. (M2) & \mybarchart{22.1pt} 66.2\% & \mybarchart{21.2pt} 63.6\% & 2.6\%\phantom{ ***} \\
I want to have an impact on society. (M3) & \mybarchart{21.5pt} 64.4\% & \mybarchart{20pt} 60.1\% & 4.3\%\phantom{ ***}  \\
\textbf{I want to help solve a societal issue. (M4)} & \mybarchart{21.2pt} 63.5\% & \mybarchart{18pt} 54.0\% & 9.5\% **\phantom{*}  \\ %
It's my hobby. (M5) & \mybarchart{21.2pt} 63.5\% & \mybarchart{23.6pt} 70.7\% & -7.2\%\phantom{ ***}  \\ 
\textbf{I can learn or improve technology skills. (M6)} & \mybarchart{17.7pt} 53.1\% & \mybarchart{24.2pt} 72.7\% & -19.6\% ***  \\ %
I want to help other contributors in the team. (M7) & \mybarchart{17.3pt} 51.8\% & \mybarchart{20.9pt} 62.6\% & -10.8\%\phantom{ ***}  \\ 
It's my job. (M8) & \mybarchart{17.3pt} 51.8\% & \mybarchart{17.7pt} 53.0\% & -1.2\%\phantom{ ***}  \\ 
\textbf{It helps me to build my portfolio and reputation for my career. (M9)} & \mybarchart{13.1pt} 39.2\% & \mybarchart{17.8pt} 53.5\% & -14.3\% *\phantom{**}  \\% p-value = 0.02996
\textbf{I need to improve this project for my work or school studies. (M10)} & \mybarchart{9.8pt} 29.3\% & \mybarchart{13.3pt} 39.9\% & -10.6\% ***  \\ %
\textbf{I want to meet new people. (M11)} & \mybarchart{5.9pt} 17.6\% & \mybarchart{10.1pt} 30.3\% & -12.7\% **\phantom{*}  \\ %
My organization/boss encourages me to work on this project. (M12) & \mybarchart{5.1pt} 15.3\% & \mybarchart{7.1pt} 21.2\% & -5.9\%\phantom{ ***}  \\
I want to get paid. (M13) & \mybarchart{4.2pt} 12.6\% & \mybarchart{4.5pt} 13.6\% & -1.0\%\phantom{ ***}  \\
\bottomrule
\end{tabular}
\begin{tablenotes}
      \small
      \item Wilcoxon rank sum test significance codes:   ‘***’ $p < 0.001$, ‘**’ $p < 0.01$, ‘*’ $p < 0.05$ 
    \end{tablenotes}
  \end{threeparttable}

\end{table*}

\subsection{\textbf{What are the motivations for contributing to OSS4SG projects? (RQ2)}}
\label{sec:RQ2}

\subsubsection{Interview}
First, we present the themes explaining the motivations for contributing to OSS4SG. Some themes we identified include contributing for an emotional fulfillment, helping those in need, and leaving their ``fingerprint on the world''. A summary of these themes is listed in Table~\ref{tab:motivation-theme}.

\subsubsection{Survey}\label{sec:RQ2-survey}
We included all the themes from Table~\ref{tab:motivation} as well as motivations included in previous studies on open source in our survey~\cite{alexander2002working,hertel2003motivation,bitzer2007intrinsic}.
To investigate if contributors are motivated in different ways for OSS4SG and OSS, we asked \textit{P-OSS4SG} to rate how important each motivation item was for them to contribute to OSS4SG, and asked \textit{P-OSS} to rate the items based on their opinions on OSS projects.
Each motivation item is a five-point Likert scale question: ``Not Important', ``Sightly Important'', ``Moderately Important'', ``Important'', and ``Very Important''.
Table~\ref{tab:motivation} lists the accumulated percentages of ``Important'' and ``Very Important'' for each motivation theme for OSS4SG and OSS respectively, and the difference between them (i.e., column \textit{Delta}). 

We found that, compared to OSS, for OSS4SG contributors it significantly less important to benefit themselves by learning or improving skills (M6, --19.6\%, $p<0.001$), building their portfolio (M9, --14.3\%, $p<0.05$) and helping with their own work or projects (M10, --10.6\%, $p<0.001$). 
On the contrary, OSS4SG contributors are significantly more motivated to solve a societal issue (M4, +9.5\%, $p<0.01$).

\mybox{Compared to OSS, OSS4SG contributors indicate they are significantly more motivated by solving societal issues rather than benefiting themselves through learning skills or building a career portfolio.}

\begin{table*}[t]\centering
  \begin{threeparttable}
\settoheight{\myboxheight}{1234567890\%}

\def\mybarchart#1{
\resizebox {#1} {\myboxheight} {%
\begin{tikzpicture}[]
\definecolor{clr1}{RGB}{99,99,99}
\definecolor{clr2}{RGB}{240,240,240}
\begin{axis}[
      axis background/.style={fill=gray!10, draw=gray!50},
      axis line style={draw=none},
      tick style={draw=none},
      ytick=\empty,
      xtick=\empty,
      ymin=0, ymax=1, %
      xmin=0, xmax=1]
\addplot [
      ybar interval=.5,
      fill=black,
      draw=none,
]
	coordinates {(1,1) (1,1)}; 
\addplot [
      ybar interval=.5,
      fill=black,
      draw=none,
]
	coordinates {(1,1) (0,1)}; 
\end{axis}%
\end{tikzpicture}%
}%
}

\caption{The responses \textit{P-OSS4SG} and \textit{P-OSS} to the question \emph{``Please rate how important the following factors are to you when you decide to work on a project''} (Q17, Q17'). %
The columns \textit{P-OSS4SG} and \textit{P-OSS} listed the accumulated percentages of responses from ``Important'' and ``Very Important'' (i.e., higher than ``Moderately Important''). \textit{Delta} list the difference of the percentages between \textit{P-OSS4SG} and \textit{P-OSS}.
Statistically significant differences are indicated with asterisks (*).  The items are ranked and numbered by the importance in \textit{P-OSS4SG}.  }
\label{tab:factor}

\begin{tabular}{rp{2cm}p{2cm}r}
\toprule
\multicolumn{1}{c}{Factors to Consider When Selecting a Project}& \multicolumn{1}{l}{\makecell{P-OSS4SG\\ (n=226)}} & \multicolumn{1}{l}{\makecell{P-OSS \\(n=202)}} & \makecell{Delta\\ (P-OSS4SG - P-OSS)}  \\
\midrule

I personally respect/care about the issue this project is trying to solve. (F1) & \mybarchart{27.9pt} 83.6\% & \mybarchart{25.2pt} 75.7\% & 7.9\% \phantom{***} \\
I like the idea of this project. (F2) & \mybarchart{27.7pt} 83.2\% & \mybarchart{26.1pt} 78.2\% & 5.0\% \phantom{***} \\ 
This project is active. (F3) & \mybarchart{24.8pt} 74.3\% & \mybarchart{24.1pt} 72.3\% & 2.0\%  \phantom{***} \\ 
The goal of this project meets some form of needs I care about. (F4) & \mybarchart{24.3pt} 73.0\% & \mybarchart{22.4pt} 67.3\% & 5.7\% \phantom{***} \\
This project is welcoming. (F5) & \mybarchart{22.6pt} 67.7\% & \mybarchart{21.4pt} 64.3\% & 3.4\% \phantom{***} \\ 
I fully understand the goal and value of this project. (F6) & \mybarchart{22pt} 65.9\% & \mybarchart{21.8pt} 65.3\% & 0.6\% \phantom{***} \\ 
\textbf{I trust the owner/organizer of this project. (F7)} & \mybarchart{21.2pt} 63.5\% & \mybarchart{14pt} 42.1\% & 21.4\% ***  \\ %
I feel confident in my skills to help with this project. (F8) & \mybarchart{20.8pt} 62.4\% & \mybarchart{21.9pt} 65.8\% & -3.4\% \phantom{***} \\
This project is well-maintained. (F9) & \mybarchart{19pt} 57.5\% & \mybarchart{21.6pt} 64.8\% & -7.3\% \phantom{***} \\
\textbf{I can learn some new skills or enhance my skills in this project. (F10)} & \mybarchart{14.6pt} 43.8\% & \mybarchart{20.8pt} 62.4\% & -18.6\% ***  \\ %
This project targets a lot of users. (F11) & \mybarchart{8.3pt} 24.8\% & \mybarchart{9.2pt} 27.7\% & -2.9\% \phantom{***} \\
This project has a diverse contributor team (e.g., gender, race, geography). (F12) & \mybarchart{5.5pt} 16.4\% & \mybarchart{6.1pt} 18.3\% & -1.9\% \phantom{***} \\
Someone else in my community is also working on this project. (F13) & \mybarchart{5.3pt} 15.9\% & \mybarchart{5.4pt} 16.3\% & -0.4\% \phantom{***} \\ 
\textbf{This project is popular in the community. (F14)} & \mybarchart{5.1pt} 15.5\% & \mybarchart{9pt} 27.2\% & -11.7\% *\phantom{**}  \\ %

\bottomrule
\end{tabular}
\begin{tablenotes}
      \small
      \item Wilcoxon rank sum test significance codes:   ‘***’ $p < 0.001$, ‘**’ $p < 0.01$, ‘*’ $p < 0.05$ 
    \end{tablenotes}
  \end{threeparttable}

\end{table*}

\subsection{\textbf{What factors do contributors consider to select an OSS4SG project? (RQ3)} } \label{sec:RQ3}
We will first discuss the factors and properties of OSS4SG projects that contributors evaluate when they select a project to contribute to, and presented the comparison with OSS projects. 
We will then discuss the search strategies contributors use to look for OSS4SG projects.
In the end, we will present the OSS4SG contributors' project preference in terms on scale of impact (spatial, temporal, and social) and owner types.
\subsubsection{Interview}
With limited space, a summary of the factors OSS4SG contributors demonstrated to consider when selecting a project in the interview is listed in Table~\ref{tab:factor}, which are the same statements we used in our survey design. 
Participants reported factors including both properties of the projects and their own personal preferences. 

Participants expressed special considerations on owners of projects, including   preferences for certain types of owners:
\quo{If it's even a charity organization I go and look at who their sponsors are.  And if it's a government I'm already like, no, it's not gonna happen.  A political party maybe.  But government is too far for me.}{P14}
We will discuss more on contributors' preference on project owners from the survey results in Section~\ref{sec:RQ3-survey}.

Participants search for OSS4SG projects in different ways. 
Some participants search for OSS4SG projects by talking to other people in meetings or tech events that are relevant to their social interests. For example:
    \quo{I can go to like a underground, political event every night and like meet activist, tech people all the time.  It's like I can just talk to people.}{P18}
Some prefer starting with project owners:
    \quo{I went to view some of the large releases that they (owners) made [\dots] and I checked through their websites also.}{P10}
Some look for projects from their personal connections:
    \quo{A friend talk about that or that other project, and you just have a feeling, hey, maybe find you to work on this.}{P15}

An interesting search strategy for OSS4SG projects is based on the social impact (see more discussion in Section~\ref{sec:RQ3-survey}):

\quo{I was reading about Coronavirus and tracking the spread of it, and found that there are no mobile apps that provide this ability.}{P6}
Participants also use regular search engines and online open source platform to search for OSS4SG (e.g., GitHub).

\subsubsection{Survey} \label{sec:RQ3-survey}
We refined all the factors of project evaluation provided in the interviews and designed 14 statements in the survey. Similarly with~\textbf{RQ2}, we asked \textit{P-OSS4SG} to rate how important each factor was when they select an OSS4SG project, and asked \textit{P-OSS} to rate the factors based on their opinions on OSS projects.
Table~\ref{tab:factor} lists the accumulated percentages of ``Important'' and ``Very Important'' for each factor for OSS4SG and OSS respectively. Column \textit{Delta} lists the importance difference between OSS4SG and OSS. 

From Table~\ref{tab:factor}, compared to OSS, OSS4SG contributors consider the importance of ``learning skills (F10)'' by 18.6\% less significantly: this result aligns with our findings on motivations in Section~\ref{sec:RQ2-survey}.  
While the popularity of a project is an important factor of evaluation for OSS, when selecting an OSS4SG project, contributors focus less on project popularity (F14, --11.7\%, $p<0.05$).
One interesting and significant difference on factors of evaluations between OSS4SG and OSS projects is the owner of projects: when selecting OSS4SG projects, contributors investigate more on the owners than OSS projects (+21.4\%, $p<0.001$). This quantitatively verifies our observations in the interviews.

In the survey, we further explored how contributors evaluated different scales of impact and owners when they select OSS4SG projects to contribute to. 

\vspace{.25em}
\noindent\textbf{Scale of Impact:} Table~\ref{tab:need} shows all respondents decisions on three pairs of project options (Project \textbf{\textit{A}} and \textbf{\textit{B}}) based on different scales of impact: spatial proximity, temporal proximity, and social proximity. 
We found that contributors prioritized OSS4SG projects that met global needs, had long-term benefits, and benefited their personal connections. 

\vspace{.25em}
\noindent\textbf{Owner Preference:} In the survey, 55.7\% of respondents expressed willingness to contribute to OSS projects hosted by technology companies, while only 44.4\% of respondents are willing to work for OSS4SG projects hosted by technology companies ($p<0.001$). There are no significant differences on preference on government (30.9\% for OSS4SG, 25.9\% for OSS) and charity (68.2\% for OSS4SG, 66.7\% for OSS) owners.

\mybox{While sharing common concerns when selecting projects, compared to OSS, OSS4SG contributors more thoroughly investigate the owners of projects. OSS4SG contributors tend to prioritize  projects  that meet  global  needs,  have  long-term  benefits,  and  benefit  their personal connections. They also expressed  less  interest  in  projects  hosted  by  technology  companies. }

\begin{table}\centering
\settoheight{\myboxheight}{1234567890\%}

\def\mybarchart#1{
\resizebox {#1} {\myboxheight} {%
\begin{tikzpicture}[]
\definecolor{clr1}{RGB}{99,99,99}
\definecolor{clr2}{RGB}{240,240,240}
\begin{axis}[
      axis background/.style={fill=gray!10, draw=gray!50},
      axis line style={draw=none},
      tick style={draw=none},
      ytick=\empty,
      xtick=\empty,
      ymin=0, ymax=1, %
      xmin=0, xmax=1]
\addplot [
      ybar interval=.5,
      fill=black,
      draw=none,
]
	coordinates {(1,1) (1,1)}; 
\addplot [
      ybar interval=.5,
      fill=black,
      draw=none,
]
	coordinates {(1,1) (0,1)}; 
\end{axis}%
\end{tikzpicture}%
}%
}

\caption{The responses to the question \emph{``Assume you only have time to work on one project, please choose the one from each of the three pairs below''} (Q21). The three pairs of projects are based on the scale of impact in terms of spatial, temporal and social proximity.
}
\label{tab:need}

\begin{tabular}{@{~}lp{2cm}@{~}}
\toprule
\multicolumn{1}{@{~}l}{Project Selection Based on Scale of Impact}& \multicolumn{1}{l@{~}}{\makecell{Percentage\\(n=404)}}  \\
\midrule
\textsc{Spatial Proximity} \\[2pt]
\makecell[l]{\textbf{A: }A project that is needed globally\\ (e.g., tracking pandemic issues like COVID19)\vspace{4pt}} & \mybarchart{32.9pt} 65.8\% \\
\makecell[l]{\textbf{B: }A project that is needed only in my local area\\ (e.g., tracking local health issue) } & \mybarchart{17.1pt} 34.2\% \\
\midrule
\textsc{Temporal Proximity} \\[2pt]
\makecell[l]{\textbf{A: }A project that is beneficial in the long term\\ (e.g., monitor global warming)}\vspace{4pt} & \mybarchart{28.85pt} 57.7\% \\
\makecell[l]{\textbf{B: }A project that is beneficial now \\(e.g., monitor a recent flood disaster)} & \mybarchart{21.1pt} 42.3\% \\
\midrule
\textsc{Social Proximity} \\[2pt]
\makecell[l]{\textbf{A: }A project that a stranger needs\\ (e.g., monitoring system for a health issue that \\ does not affect my family)\vspace{4pt}} & \mybarchart{16pt} 31.9\% \\
\makecell[l]{\textbf{B: }A project that someone I know personally needs \\(e.g., diabetes tracker my family, friend or myself \\ can use) } & \mybarchart{34pt} 68.1\% \\
\bottomrule
\end{tabular}

\end{table}

\begin{table*}[t]\centering
\caption{The responses from \textit{P-OSS4SG+} and \textit{P-OSS+} to the question \emph{``Please rate the agreement on how challenging the following aspects are to you''} (Q27, Q27'). 
The columns \textit{P-OSS4SG+} and \textit{P-OSS+} listed the accumulated percentages of responses from ``Agree'' and "Strongly Agree" (i.e., higher than ``Neither Agree nor Disagree''). \textit{Delta} listed the difference of the percentages between \textit{P-OSS4SG+} and \textit{P-OSS+}. Statistically significant differences are indicated with asterisks (*). The challenge items are ranked and numbered by descending agreement of \textit{P-OSS4SG+}.  }
\label{tab:challenge}

  \begin{threeparttable}
\settoheight{\myboxheight}{1234567890\%}

\def\mybarchart#1{
\resizebox {#1} {\myboxheight} {%
\begin{tikzpicture}[]
\definecolor{clr1}{RGB}{99,99,99}
\definecolor{clr2}{RGB}{240,240,240}
\begin{axis}[
      axis background/.style={fill=gray!10, draw=gray!50},
      axis line style={draw=none},
      tick style={draw=none},
      ytick=\empty,
      xtick=\empty,
      ymin=0, ymax=1, %
      xmin=0, xmax=1]
\addplot [
      ybar interval=.5,
      fill=black,
      draw=none,
]
	coordinates {(1,1) (1,1)}; 
\addplot [
      ybar interval=.5,
      fill=black,
      draw=none,
]
	coordinates {(1,1) (0,1)}; 
\end{axis}%
\end{tikzpicture}%
}%
}

\begin{tabular}{rp{2cm}p{2cm}c}
\toprule
\multicolumn{1}{c}{Challenges}& \multicolumn{1}{l}{\makecell{P-OSS4SG+\\ (n=153)}} & \multicolumn{1}{l}{\makecell{P-OSS+ \\(n=228)}} & \makecell{Delta\\ (P-OSS4SG+ - P-OSS+)}  \\
\midrule
It is hard for newcomers to understand how to contribute to the project. (C1) & \mybarchart{23.3pt} 69.9\% & \mybarchart{21.5pt} 64.5\% & 4.5\% \phantom{**} \\
It is hard to understand what features my users need. (C2) & \mybarchart{17.9pt} 53.6\% & \mybarchart{16.4pt} 49.1\% & 4.5\% \phantom{**} \\
Not knowing where to find good projects to work on. (C3) & \mybarchart{16.6pt} 49.7\% & \mybarchart{12.3pt} 36.8\% & 12.9\% \phantom{**} \\ 
Needing more money to work on a project. (C4) & \mybarchart{16.3pt} 49.0\% & \mybarchart{14.3pt} 43.0\% & 6.0\% \phantom{**} \\ 

Not understanding the direction of a project. (C5) & \mybarchart{13.9pt} 41.8\% & \mybarchart{11.4pt} 34.2\% & 7.6\% \phantom{**} \\
Other contributors losing sight of direction of a project. (C6) & \mybarchart{10.7pt} 32.0\% & \mybarchart{11.8pt} 35.5\% & -3.5\% \phantom{**} \\
Working with people who do not understand open source. (C7) & \mybarchart{10.7pt} 32.0\% & \mybarchart{11.3pt} 33.8\% & -1.8\% \phantom{**} \\ 
The project is over-engineered. (C8) & \mybarchart{8.3pt} 24.8\% & \mybarchart{8.2pt} 24.6\% & 0.2\% \phantom{**} \\ 
\textbf{Stakeholders are unreasonable on feature requests. (C9)} & \mybarchart{7.4pt} 22.2\% & \mybarchart{9.8pt} 29.4\% & -7.2\% ** \\ %
\textbf{Too much time is spent on documentation in this project. (C10)} & \mybarchart{4.8pt} 14.4\% & \mybarchart{7pt} 21.1\% & -6.7\% *\phantom{*} \\ %

\bottomrule
\end{tabular}
\begin{tablenotes}
      \small
      \item Wilcoxon rank sum test significance codes:   ‘***’ $p < 0.001$, ‘**’ $p < 0.01$, ‘*’ $p < 0.05$ 
    \end{tablenotes}
  \end{threeparttable}

\end{table*}

\subsection{\textbf{What are the current challenges to work for OSS4SG? (RQ4)} }\label{sec:RQ4}
We report the challenges contributors are facing with currently in OSS4SG. In the end of the section, we discuss on observations from P21, who provided insightful and unique feedback on challenges in OSS4SG from the perspectives of technical advisors (with no programming background).

\subsubsection{Interview}
The most frequently reported challenge in our interviews is that it is very hard to match contributors and OSS4SG projects, (P1, P3, P4, P9, P18). This challenge is also indicated from consulting with the GitHub Social Impact Sector. Currently, there is no indicator (e.g., badges) yet to recommend or identify OSS4SG projects in open source community:
    \quo{It is difficult to know where the projects are.  Where the communities are.  And getting involved in it.  There are many, many, many developers that might want to contribute, but they never get, you know, an announce or publication, a post, something.}{P1}

Lack of funding, which results in unstable flow of contributors, is reported to be a challenge for OSS4SG:
    \quo{I honestly think the hardest thing about working on social good is very frequently they're funded by charities, so it's very hard to get people's full focus on it.  Like, paid full focus on it.}{P4}

Because OSS4SG projects often include contributors with different backgrounds, communication can be challenging:
    \quo{In a social good project, we would also have engineers and designers, so it’ll take a lot of time to come to a conclusion because there are so many different skill perspectives on the table.}{P17}

The lack of concept of open source of users make it challenging for OSS4SG controbutors, emotionally:
     \quo{For developers maybe it's quite common, but for normal people, it's not familiar with that concept [of open source], like that you can contribute something back and that is free, so you like blame the developer for error or something\dots}{P3}

Participants also reported conflicts they have experienced in OSS4SG. For example, some contributors may be ignorant of a wider requirement of usage:
    \quo{[Talking about other contributors] So, we can use the software this way, you know, for our good.  I don’t care the others.}{P1}

Different cultural backgrounds may introduce conflicts on social norms:

    \quo{How they express, it would be much frank.  It could be, at times, by putting their opinions, they might come off to some people as harsh.}{P17}

Different backgrounds (especially working with contributors with less technology background) may introduce conflicts on technology-based communication:

    \quo{I sometimes get told by them that I'm speaking maybe on a too technical of level.}{P14}

\subsubsection{Survey}
In our survey, we grouped and refined all the observations on challenges in OSS4SG in the interviews, and then combined with challenges in OSS from previous studies~\cite{lee2017understanding}. Finally we designed 10 challenge statements.
We asked \textit{P-OSS4SG+} (i.e., with actual OSS4SG experience) to rate how much they agreed on each challenge in OSS4SG, and asked \textit{P-OSS+} to rate the agreement on each challenge based on their experience in OSS.
Each challenge statement is a five-point Likert scale question: ``Strongly Disagree'', ``Disagree'', ``Neither Agree nor Disagree'', ``Agree'', and ``Strongly Agree''.
Table~\ref{tab:challenge} list the accumulated percentages of ``Agree'' and ``Strongly Agree'' for each challenge in OSS4SG and OSS respectively. This table rank all the challenges by the agreement percentage of OSS4SG.

From Table~\ref{tab:challenge}, for both OSS4SG and OSS, ``It is hard for newcomers to understand how to contribute to the project. (C1)'' is reported to be the most challenging factor. OSS4SG contributors reported a higher rate on ``not knowing where to find good projects to work on (C3)'' ($\textit{Delta} = 12.9\%$), which quantitatively verified the observations in the interviews. With a significant difference on agreement, OSS4SG contributors face less challenges with unreasonable feature requests (C9) and over-documentation (C10).
In general, OSS4SG and OSS contributors face similar challenges except that it is more difficult for OSS4SG contributors to find projects to work on.

\subsubsection{Interview with a technical advisor}
We had one participant, P21, from our interviews who was not a software developer but had been working for years as a technical advisor that coordinated between software development teams and stakeholders (in healthcare and pharmacy) to convert their products to OSS. 
The perspective he provided highlights the value of community-centric and project-centric roles~\cite{trinkenreich2020hidden}.
He provided insightful feedback from years of experience working on OSS4SG that involved collaborating with multiple local governments and administrations globally:
\quo{It takes a long time to set up the global communities around the software...
The biggest challenge is to get a political consensus in the given countries that this is what we are going to do and formalize it and making sure that you have national level consensus and a plan, commitment.}{P21}

He also pointed out that, for an OSS4SG project that may need the support of political administrations, it was necessary to consider the potential reachability and limitations in the project's design phase. Furthermore, he also suggested that the OSS4SG community needed a formal model to convert social good software projects to open source. Currently, they could only explore potential options by learning from individual cases:
\quo{So what we have been looking at as a model is [elided software name], or the open source application for district health information system. Because I think they have done a very good job in regards to both publishing the software, but also, creating a community around it that, of course, massively increases sustainability.}{P21}

Though P21 does not have programming experience, his experience on collaborating with non-technology stakeholders and pushing social good products to OSS is valuable for OSS4SG community.

\mybox{Though facing with similar challenges with OSS, it is more challenging for OSS4SG contributors to find a project to contribute to.}

\section{Discussions and Implications}
Our findings demonstrate that \OSSFSG is a sub-type of OSS projects that have its own characteristics. 
Developers perceive differences between OSS4SG and general OSS on definition, motivation, factors used for evaluation, and challenges.
In the remainder of this section, we present design implications based on our findings specifically for OSS4SG contributors, project owners, organizations seeking for OSS4SG options, and fund raisers to improve the OSS4SG community.

\subsection{Match contributors to OSS4SG projects} 
In our study, OSS4SG contributors described challenges with the lack of information and communication channels to discover OSS4SG projects. They also indicated that OSS4SG projects that need help cannot effectively reach developers willing to help (Section~\ref{sec:RQ4}).
Thus, we suggest implications to mitigate these challenges from the perspective of the OSS community and project owners.

For the \emph{OSS community}, we suggest \textbf{instrumenting badges and labels} and \textbf{improving nomination guidelines} to highlight OSS4SG projects. 
Research has shown that indicating OSS project properties with badges can improve the participation and help contributors search for projects of interest~\cite{trockman2018adding}. 
We suggest a similar approach to help contributors identify projects that are for social good. Our findings suggest contributors identify OSS4SG projects based on the targeted users and social issues. Such badges or labels can be added to OSS projects to ease the searching process. Currently, third parties that provide directories of OSS4SG projects (e.g., Ovio, Digital Public Goods, etc.) use free nominations from users. We suggest curators of OSS4SG project directories adapt nomination guidelines with criteria of OSS4SG to lead users to effective nominations.

For \emph{OSS4SG advocates, project owners, organizers, and fund raisers}, we suggest clarifying and \textbf{emphasizing relevant OSS4SG information to attract contributors of interest}. 
Our findings on OSS4SG motivations (Section~\ref{sec:RQ2}) and factors used for selecting projects (Section~\ref{sec:RQ3}) indicate that OSS4SG contributors care more about project owners, projects' goals, social impact and targeted users more than when compared to general OSS contributors. 
Emphasizing this information in project documentation or websites can attract and help contributors with interest of certain type of projects. 
Thorough introductions and easily accessible links to websites of project owners and organizations are also helpful for contributors to evaluate and select projects.

\subsection{Protect safety and privacy in OSS4SG}
Our findings suggest that many OSS4SG projects focus on particularly sensitive societal issues and target empowering marginalized populations of users. Therefore special care may be required to support contributors and end users in OSS4SG effectively and respectfully. 

We suggest providing required training and reporting structures for disruptive behaviors in OSS4SG projects to \textbf{protect contributors and end users when risks may apply.}
For example, OSS4SG projects that serve victims of domestic violence (P4's project), or gender minorities (P14's project), may involve interactions with users, in which case both contributors and end users may suffer from polarized opinions, even harm, especially in certain geographical regions. 
OSS4SG contributors may also lack the knowledge and  or not recognize signals to potentially risky situations. Thus, extra support of reporting and training is necessary.
 
In OSS4SG projects that collaborate with non-traditional entities, such as local governments and charities, or involve sensitive data, such as COVID-19 tracking systems, \textbf{special regulations may apply to protect data privacy} and be compliant with government regulations and policies. In some cases, these regulations may be communicated by stakeholders (P21). OSS4SG contributors and project owners should be aware of regulations and be equipped to resolve potentially harmful outcomes from violations of data privacy.

\section{Limitations}

In this section we discuss the limitations and threats to validity of this study.  
The observations and themes that are discussed in this paper are based on 21 interviews and 517 survey responses within the open source community. As with any empirical study the extent to which the findings generalize to other populations is unknown; however, we are confident that the results from this apply to the broad open source community. Our sampling approach for the interviews was specifically designed to capture a wide range of responses~\cite{patton1990qualitative} by using a stratified sample based on project popularity, contribution count, and commit recency. The participants for the interviews and surveys were selected from 437 manually \emph{curated} OSS4SG projects, which cover most of the 17 sustainable development goals from the United Nations~\cite{sdg}. 

The survey compared OSS4SG experiences with experiences in regular OSS projects. Since there is no automated way to decide whether a OSS project is for social good, we relied on self-reported information from survey participants to distinguish the groups. There are limitations with any survey, for example, since survey participants self-selected, contributors to social good projects may have been more eager to respond. Participants may also have misunderstood some questions; to reduce this risk the survey was reviewed by several people who are not co-author of the paper and piloted with a small group. As the survey was in English, only English-speakers participated. Therefore we cannot claim that the results apply to regions with less English-proficiency. Investigating how the definitions of social good and motivations of social good vary across regions and racial/ethnic minorities is a promising direction for future research. 

\vspace{-.3em}
\section{Related Work}
\subsection{Social Good outside Computer Science} %

The term~\textit{Social Good} has been embraced in domains such as sociology and social work~\cite{mor2019social,mor2020practice}, business and non-profit world~\cite{gao2014commitment,mcwilliams2006corporate,quayson2020technology}.
Mor Barak refers to social good as the heart of the social work profession~\cite{mor2020practice} and pointed out that, in contrast with  \textit{public goods}, social good does not depend on public policy and funding. Similarly, social good should be distinguished from \textit{common good} which is tied to a specific community's goals, norms, and resources~\cite{mor2019social}.
Social good has been essential to promote activities and attract public attention and support: hackathons evolved in social work with new goals of changing the world for good~\cite{henderson2015getting}, and contributing to social good can improve volunteers' happiness and well-being~\cite{borgonovi2008doing,wilson2003doing}.
In management science, social good is discussed in the context of corporation social responsibility with the goal to positively impact both corporations and society~\cite{gao2014commitment,mcwilliams2006corporate,quayson2020technology}.

\subsection{Social Good in CS Education, Security, AI, and SE}\label{sec:related-work-se}

In the computer science domain, researchers and practitioners in CS education have been enhancing underrepresented groups' participation in computational activities to promote diversity and education equity~\cite{robinson2014underrepresented, margolis2012beyond,payton2016stars,mejias2018culturally, washington2019ethnic, mcfarlane2020get, ong2011inside, braswell2020pivoting}.
Security researchers introduced social good through emphasizing the importance of data privacy in tandem with cyber attacks and a focus on citizen rights and safety~~\cite{karafili2017argumentation, bandara2009using}.
In Artificial Intelligence (AI), social good is referred to as the guarantee for appropriate moral choices~\cite{abebe2018mechanism}, in particular trust, fairness, and identification of bias in AI and machine learning models~\cite{varshney2019trustworthy, wearn2019responsible, binns2018fairness, mehrabi2019survey}. 
The \emph{AI for Social Good} (AI4SG) movement leverages AI techniques for social good and several technology companies started initiatives to tackle social problems and improve lives using AI-based solutions~\cite{cowls2019designing, berendt2019ai} Examples are Microsoft's AI for Earth~\cite{ai4earth} and Google's AI for Social Good~\cite{ai4good}.

In software engineering, Ferrario et al.~\cite{ferrario2014software} referred to ``software engineering for social good'' as  the  development,  maintenance  and  sustainability  of software aimed to promote social change. They proposed a process to allow social software projects to be built quickly with limited resources, while helping the understanding of the social context and vulnerable user group.
Dekhtyar et al.~\cite{dekhtyarre2020} suggested that SE conferences should be the focal points for activities that benefit society and that such activities should be an integral, valued, and recognized part of the conference programs. They shared their experience of running such an event (RE Cares) with great success~\cite{dekhtyar2019requirements}.

In April 2020, GitHub's Social Impact Sector released a report indicating the tremendously increasing demands and lack of research and resources related to open source projects with social impact~\cite{github2020socialsectorreport}. Although the aforementioned report presents discussions from the organization and fundraiser perspective, it does not primarily focus on software developers.
To the best of our knowledge, our paper introduces the notion of OSS4SG and presents a first empirical study of social good in the open source community from the software developer perspective. We hope this work can lead to future community efforts to support OSS4SG more efficiently.

\subsection{Social Factors in Open Source}

Studies have investigated fundamental features and effective work practice for OSS contributors~\cite{ghosh2002free, crowston2004effective, haefliger2008code, von2003open}.
Hars et al. studied motivations of open-source contributors in technology and found that contributors are most concerned with self-marketing and fulfilling personal software needs~\cite{alexander2002working}. Similar findings are verified in~\cite{hertel2003motivation,bitzer2007intrinsic}. Our results highlight that the motivations for OSS4SG contributors are different: compared to OSS projects, contributors significantly care less about benefiting themselves.

Other studies in OSS have focused on social aspects of the software development, for example, the socialization of new contributors, that is, how 
software developers are sustained and reproduced over time through the progressive integration of new members in OSS projects~\cite{ducheneaut2005socialization}.
OSS contributors preferentially join projects when they have pre-existing social relationships~\cite{casalnuovo2015developer}. Their participation in social networks may have direct impact on OSS success~\cite{fangneed}.
Research has also looked at donation patterns in OSS~\cite{overney2020not} and investigated the challenges and support for underrepresented participant groups~\cite{vasilescu2015gender, padala2020gender}. Gender and tenure diversity was found to present a positive effect on productivity~\cite{vasilescu2015gender}. However, OSS suffers from bias against underrepresented groups (e.g., women)~\cite{padala2020gender, terrell2017gender}.
Qiu et al. showed that OSS contributors used signals, such as the amount of recent activities, to select which projects to work on~\cite{qiu2019signals}. Likewise, Ford et al. identified how contributors used those signals to evaluate each other~\cite{ford2019beyond}.
Our findings on project selection (Section~\ref{sec:RQ3-survey}) are similar to regular OSS; however, when selecting projects, OSS4SG contributors paid more attention to the project owners than OSS contributors. 

While a rich body of research in OSS sheds light on how to better support the community with the understanding of effects of different social factors, the findings may overlook unique properties of OSS4SG. As one of the first steps, our study compensates the investigation for OSS4SG to support future research, service and improvement for this community.

\vspace{-.3em} 
\section{Conclusion}
Open Source is an important distributed and collaborative development platform for software engineering.
Social good projects in open source have demonstrated prevalent values to both software engineering and society, which has not yet been explored in the software engineering research community. 
In this paper, we introduced the notation of \textit{Open Source Software for Social Good (OSS4SG)} to the community. We presented the first empirical study based on 17 hours of semi-structured interviews and 517 survey responses of motivations and challenges in OSS4SG. %
We believe our results shed light on the global impacts open source software can have and will lead to future research and support for OSS4SG, including but not limited to, providing help and steering guidance for adapting more social good projects to OSS.  

Upon the time of publishing, this work has been featured in the {2020 \gh Octoverse Community Report}~\cite{github2020octoversecommunity}. In our future work we plan to actualize our findings and devise interventions to support contributors identifying projects and project owners identifying the contributors they need to be successful.

 \section*{Acknowledgment}

We want to extend gratitude to our study participants for providing such elaborate and personal narratives of what motivated them to be innovators. We would also like to thank Mala Kumar from \gh's Tech for Social Good program of the Social Impact Team for providing insight in early stages of this project. 
The values and implications of this work to open source software for social good have been shared with the \gh Social Impact Team to promote community growth and support. 
The ethics for this study were reviewed and approved by the Microsoft Research Institutional Review Board (MSRIRB), which is an IRB federally registered with the United States Department of Health \& Human Services. (Reference: MSRIRB \#649 and \#718).

\bibliographystyle{IEEEtran}
\bibliography{mybib}

\end{document}